\documentclass[iop,revtex4]{emulateapj}
\usepackage{graphics,graphicx}
\usepackage{helvet}
\usepackage[utf8]{inputenc}
\usepackage[T1]{fontenc}
\usepackage{longtable}
\usepackage{soul}
\usepackage{hyperref}
\usepackage{amsmath, amssymb, gensymb}
\usepackage{ulem}
\usepackage{natbib}
\usepackage{microtype}
\usepackage{multirow}
\usepackage{morefloats}
\usepackage{rotating}
\usepackage{xspace}
\def\kms {km\,s$^{-1}$\xspace}

\def\jybmkms {Jy\,bm$^{-1}$\,km\,s$^{-1}$\xspace}
\def\mjybmkms {mJy\,bm$^{-1}$\,km\,s$^{-1}$\xspace}

\def\mjybm {mJy\,bm$^{-1}$\xspace}
\def\ujybm {$\mu$Jy\,bm$^{-1}$\xspace}

\def\jybmkms {Jy\,bm$^{-1}$\,km\,s$^{-1}$\xspace}
\def\ergpersec {erg\,s$^{-1}$\xspace}
\def\ergpersecpercmsq {erg\,s$^{-1}$\,cm$^{-2}$\xspace}
\def\vlsr {$v_{\textrm{LSR}}$\xspace}
\def\etal {\textit{et al.}\xspace}
\newcommand{\mir}[1]{\textbf{\fontfamily{lmvtt}\selectfont #1}} % font for code, e.g., MIRIAD

\def\co {CO($J$\,=\,2\,$\rightarrow\,$1)\xspace}

\begin{document}

% Plot directory
\graphicspath{ {figures/} }

\slugcomment{\textit{Accepted for publication in ApJL on 12 Apr 2016}}

\title{An extremely high velocity molecular jet surrounded by an ionized cavity  \\ in the protostellar source Serpens SMM1}
\shorttitle{High-velocity CO jet and ionized outflow cavity in Serpens SMM1}

\author{Charles~L.~H.~Hull,\altaffilmark{1,2} Josep~M.~Girart,\altaffilmark{1,3} Lars~E.~Kristensen,\altaffilmark{1} Michael~M.~Dunham,\altaffilmark{1} Adriana Rodr\'iguez-Kamenetzky,\altaffilmark{4,5} Carlos~Carrasco-Gonz\'alez,\altaffilmark{5} Paulo~C.~Cort\'es,\altaffilmark{8,9} Zhi-Yun~Li,\altaffilmark{7} \\ and Richard~L.~Plambeck\altaffilmark{6}}

\shortauthors{Hull \etal}
\email{chat.hull@cfa.harvard.edu}

\altaffiltext{1}{Harvard-Smithsonian Center for Astrophysics, 60 Garden St., Cambridge, MA 02138, USA}
\altaffiltext{2}{Jansky Fellow of the National Radio Astronomy Observatory}
\altaffiltext{3}{Institut de Ci\`encies de l'Espai, (CSIC-IEEC), Campus UAB, Carrer de Can Magrans S/N, 08193 Cerdanyola del Vall\`es, Catalonia, Spain}
\altaffiltext{4}{Instituto de Astronom\'ia Te\'orica y Experimental, (IATE-UNC), X5000BGR C\'ordoba, Argentina}
\altaffiltext{5}{Instituto de Radioastronom\'ia y Astrof\'isica (IRyA-UNAM), 58089 Morelia, M\'exico}
\altaffiltext{6}{Astronomy Department \& Radio Astronomy Laboratory, University of California, Berkeley, CA 94720-3411, USA}
\altaffiltext{7}{Astronomy Department, University of Virginia, Charlottesville, VA 22904, USA}
\altaffiltext{8}{Joint ALMA Observatory, Av. Alonso de C\'ordova, 3104 Santiago, Chile}
\altaffiltext{9}{National Radio Astronomy Observatory, 520 Edgemont Rd, Charlottesville, VA 22903, USA}

\begin{abstract}
\noindent
We report ALMA observations of a one-sided, high-velocity ($\sim$\,80\,\kms{}) \co jet powered by the intermediate-mass protostellar source Serpens SMM1-a.  The highly collimated molecular jet is flanked at the base by a wide-angle cavity; the walls of the cavity can be seen in both 4\,cm free-free emission detected by the VLA and 1.3\,mm thermal dust emission detected by ALMA.  This is the first time that ionization of an outflow cavity has been directly detected via free-free emission in a very young, embedded Class 0 protostellar source that is still powering a molecular jet.
The cavity walls are ionized either by UV photons escaping from the accreting protostellar source, or by the precessing molecular jet impacting the walls. 
These observations suggest that ionized outflow cavities may be common in Class 0 protostellar sources, shedding further light on the radiation, outflow, and jet environments in the youngest, most embedded forming stars.
\\
\end{abstract}


\section{INTRODUCTION}
\label{sec:intro} 

There are several examples of extremely high-velocity (EHV), highly collimated, $\gtrsim$\,100\,\kms{} molecular $^{12}$CO and SiO jets that are driven by protostellar sources \citep{Guilloteau1992, Tafalla2004, Hirano2010}.  Some of these jets, such as those in the Class 0 protostar L1448C, exhibit proper motions suggesting velocities of nearly 200\,\kms{} \citep{GirartAcord2001}.  These fast molecular jets only appear in the very youngest, Class 0 protostellar objects, suggesting that molecular jets are short-lived.
%as protostars age their molecular outflows tend to become wider and slower \citep{ArceSargent2006}.
However, atomic jets with velocities of several $\times$ 100\,\kms{} can be seen after the Class 0 phase in optical and infrared tracers toward evolved T Tauri stars such as HL Tau and XZ Tau \citep{Mundt1990, EisloffelMundt1998}.  Many low- and high-mass protostellar jets show free-free emission tracing internal shocks \citep{Marti1993, Rodriguez2000} and even synchrotron emission from shocks against a dense ambient medium \citep{CarrascoGonzalez2010, RodriguezKamenetzky2016}.

One source with a powerful free-free radio jet is Serpens SMM1, the brightest millimeter-wavelength source in the Serpens Main star-forming region ($d = 415$\,pc; \citealt{Dzib2010}), hereafter known as ``SMM1.''\footnote{SMM1 has been known by many names including Serpens FIRS1, Serp-FIR1, Ser-emb 6, IRAS 18273+0113, S68 FIR, S68 FIRS1, and S68-1b.}  SMM1 is an intermediate-mass Class 0 protostellar source that harbors a strikingly massive ($\sim$\,1\,$M_{\odot}$) disk \citep{Hogerheijde1999, Enoch2009}.
%The star-forming dark cloud near the Sharpless 68 (S68) optical nebula in Serpens has been studied extensively at centimeter wavelengths, first by \citet{Rodriguez1980}, who were studying radio sources in the vicinity of Herbig-Haro (HH) objects.  

The radio jet powered by SMM1, known as the ``Serpens triple radio source,'' was first discovered by \citet{Rodriguez1980}, who identified two compact 5\,GHz radio sources; \citet{Snell1986} later identified three.  These radio sources were later described as a radio jet due in part to the large, symmetric proper motions of the NW and SE radio sources \citep{Rodriguez1989, Curiel1993, RodriguezKamenetzky2016}.  \citet{RodriguezKamenetzky2016} discuss the presence of synchrotron emission from shocks against the dense, ambient molecular material.
%most of these jets have spectra consistent with free-free emission, which is true of the central source in the VLA map coincident with the ALMA dust continuum peak (see Figure \ref{fig:ALMA_VLA}, right panel).

We report observations from the Atacama Large Millimeter-submillimeter Array (ALMA) that show both the dust continuum and the outflow activity in the region with unprecedented clarity.   We also show centimeter-wave data from the Karl G. Jansky Very Large Array (VLA), reported in \citet{RodriguezKamenetzky2016}, that show an ionized cavity surrounding a highly collimated molecular jet.

\begin{figure*} [hbt!]
\centering
\includegraphics[scale=0.39, clip, trim=0cm 0cm 0cm 0cm]{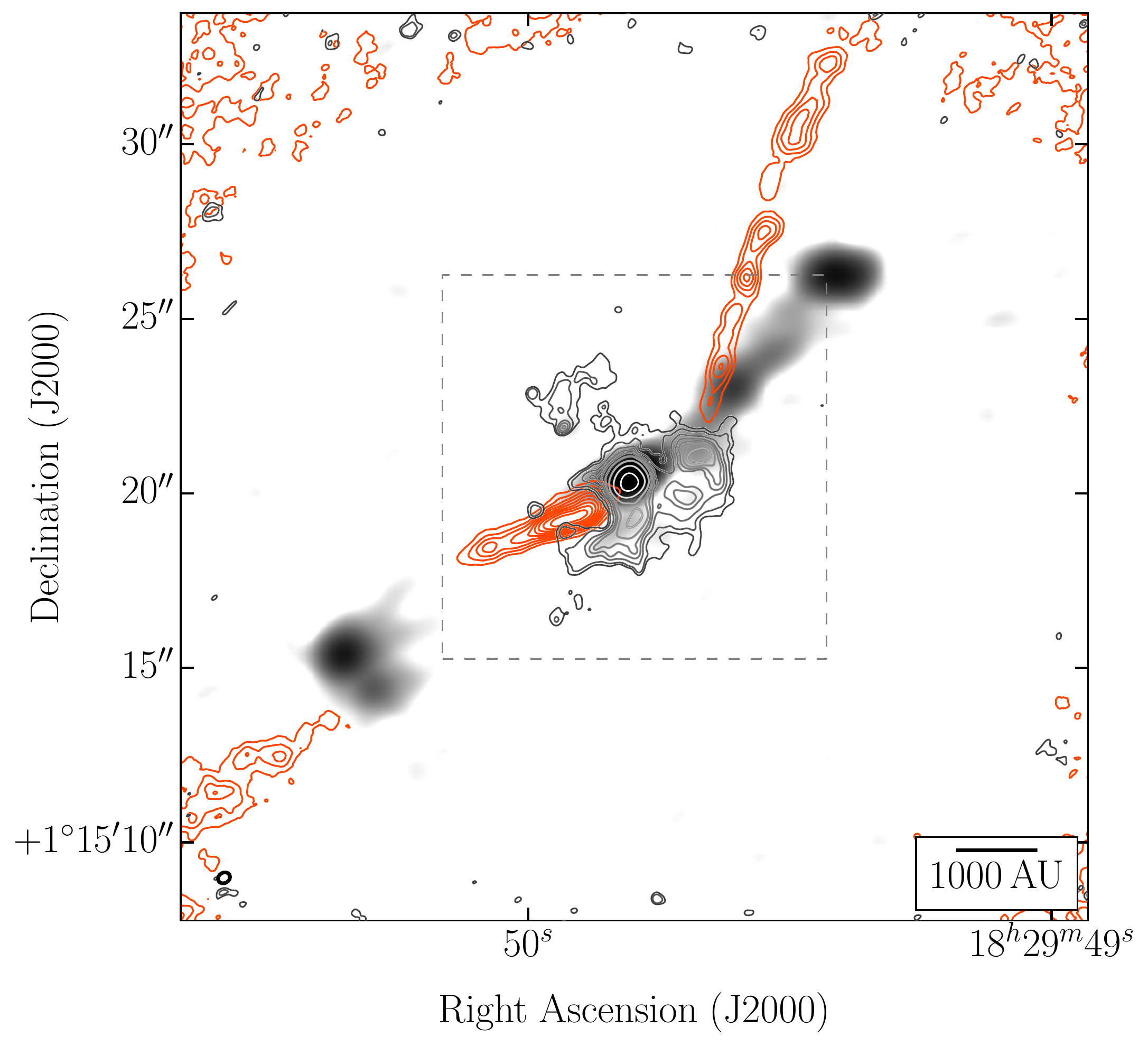}
\includegraphics[scale=0.39, clip, trim=0cm 0cm 0cm 0cm]{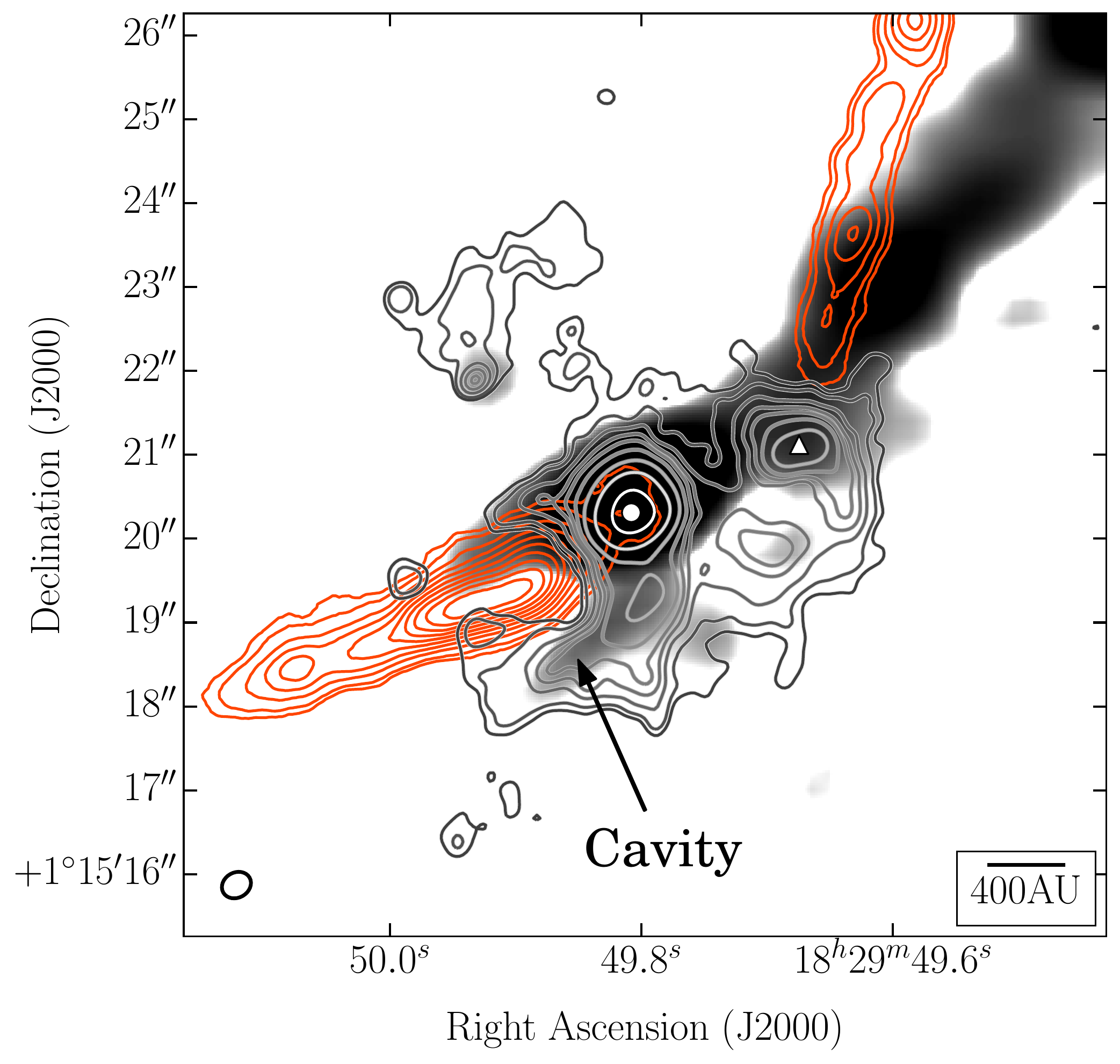}
\caption{\footnotesize   
Maps of ALMA 1.3\,mm dust continuum (contours with gradation), VLA 4\,cm image (grayscale), and extremely high-velocity \co{} emission (red contours; there is no blueshifted EHV emission).  
Velocities are relative to SMM1's \vlsr{} = 9\,\kms{}.  
The images are centered on the main source SMM1-a (white circle: $\alpha_{\textrm{J2000}} =$ 18:29:49.81, $\delta_{\textrm{J2000}} =$ +1:15:20.32); the fainter source $\sim$\,2$\arcsec$ to the WNW is SMM1-b (white triangle: $\alpha_{\textrm{J2000}} =$ 18:29:49.67, $\delta_{\textrm{J2000}} =$ +1:15:21.07).
\textit{Left:} The dashed gray zoom box represents the region pictured in the right-hand panel.
%The noise in the CO map increases toward the edges of the map because of the primary beam correction.
\textit{Right:} The moment 0 CO map comprises integrated emission from 40--80\,\kms{}; contours are --3, 3, 4, 6, 8, 10, 12, 14, 16, 18, 20 $\times$ the rms noise level in the map of 30\,\mjybmkms{} (the positive CO contours in the left panel start at 4 $\times$ the rms noise).
The 1.3\,mm ALMA dust contours are --3, 3, 4, 6, 8, 10, 14, 26, 74, 266 $\times$ the rms noise level of 0.5\,\mjybm{}.
The VLA 4\,cm emission peaks at 1.55\,\mjybm{} and has an rms noise level of 6\,\ujybm{}.
As marked by the arrow and label, both the ALMA and VLA emission trace a cavity to the SE of the source; the CO jet emanates from within the cavity.  
} 
\vspace{0.2in}
\label{fig:ALMA_VLA} 
\end{figure*}

\section{OBSERVATIONS}

The 1.3\,mm (Band 6) ALMA data we present include wide-band dust continuum windows as well as narrow-band spectral windows including the \co{} transition.
%\sio{}, \cs{}, \formaldehyde{}, \ceighteeno{}, and \dco{} transitions.  
%Here we will discuss only the continuum and high-velocity CO results; a more thorough treatment of the other species will be left for a future paper.  
The 1.35\,GHz of dust continuum bandwidth comprise both dedicated continuum windows as well as the portions of spectral line windows that were uncontaminated by line emission; the windows range in frequency from $\sim$\,216--232\,GHz.  The dust continuum image, most clearly seen in Figure \ref{fig:ALMA_VLA} (right), was produced by using the CASA task \mir{CLEAN} to combine the data from the two configurations using multi-scale \mir{CLEAN} with a weighting parameter of robust\,=\,0.  The rms noise level in the dust map is 0.5\,\mjybm{}, and the final synthesized beam is $0\farcs36\times0\farcs30$ at a position angle of --60$\degree$.

The ALMA observations were taken in two different array configurations\footnote{The definition of the visibility weights changed between the two observations in 2014 and 2015.  We carefully inspected the weights from both datasets before combining the data to ensure that they were applied properly.} on 2014 August 18 ($\sim$\,0.3$\arcsec$ angular resolution) and 2015 April 06 ($\sim$\,1$\arcsec$ resolution).  
%The observations cycled through three target sources, all of which are Class 0 protostellar cores in the Serpens Main star-forming region at a distance $d = 415$\,pc \citep{Dzib2011}: Serpens-Emb\,6 (SMM1), Emb\,8, and Emb\,8(N).  
As can be seen in Figure \ref{fig:ALMA_VLA}, there are two main sources detected in the ALMA maps (\citealt{Choi2009} first suggested that SMM1 was a multiple system).  Following \citet{Choi2009} and \citet{Dionatos2014} we will refer to the brighter eastern source as SMM1-a ($\alpha_{\textrm{J2000}} =$ 18:29:49.81, $\delta_{\textrm{J2000}} =$ +1:15:20.41) and the fainter source $\sim$\,2$\arcsec$ to the WNW as SMM1-b ($\alpha_{\textrm{J2000}} =$ 18:29:49.67, $\delta_{\textrm{J2000}} =$ +1:15:21.15).   

% Coordinates from first submission: ($\alpha_{\textrm{J2000}} =$ 18:29:49.81, $\delta_{\textrm{J2000}} =$ +1:15:20.32) and the fainter source $\sim$\,2$\arcsec$ to the WNW as SMM1-b ($\alpha_{\textrm{J2000}} =$ 18:29:49.67, $\delta_{\textrm{J2000}} =$ +1:15:21.07).   

% The phase center of the ALMA images was matched to that of the VLA images: $\alpha_{\textrm{J2000}} =$ 18:29:49.795, $\delta_{\textrm{J2000}} =$ +1:15:20.790.}

% Total bandwidth: 468750.0 + 2*234375.000 + 5*58593.750 + 117187.500

The \co{} window had a velocity resolution of 0.6\,\kms{}; the center of the window was Doppler-corrected to be the line's rest frequency of 230.538\,GHz.  The CO maps are also from the combined data and were produced with a weighting parameter of robust\,=\,2, resulting in a synthesized beam of $0\farcs55\times0\farcs45$ at a position angle of --53$\degree$

The combined VLA C (4.5--6.5\,GHz) and X (8--10\,GHz) band data have an average frequency of 7.44\,GHz, corresponding to a wavelength of 4\,cm (see \citealt{RodriguezKamenetzky2016} for details).

\vspace{0.3in}
\section{RESULTS}

Our results comprise the highest-resolution, most sensitive millimeter-wave observations of SMM1 to date.  From the perspective of this work, the most striking feature is the wide-angle cavity---to the SE of SMM1-a, shaped like a backwards ``C,'' and with a $\sim$\,2$\arcsec$ diameter---detected both at 4\,cm (VLA) and at 1.3\,mm (ALMA).  Emanating from within the cavity is a one-sided, redshifted EHV CO jet ($\sim$\,30--80\,\kms{} relative to SMM1's \vlsr{} = 9\,\kms{}) driven by SMM1-a.  In Figure \ref{fig:ALMA_VLA}, a separate EHV jet can be seen in \co{} emission extending to the NNW; our data confirm definitively that this jet emanates from SMM1-b.  

The jets from SMM1-a and SMM1-b are both monopolar (and redshifted), presumably because of the geometry of the surrounding cloud.  Monopolar outflows have been seen before (see, e.g., \citealt{Loinard2013, Kristensen2013a, Codella2014}), but their origin is still unclear.  Furthermore, we see no EHV emission from the blueshifted outflow counterpart on the other (NW) side of the source, where conversely there is strong free-free emission.  We also attribute this to intrinsic source asymmetry.

\begin{figure} [hbt!]
\centering
\includegraphics[scale=0.39, clip, trim=0cm 0cm 0cm 0cm]{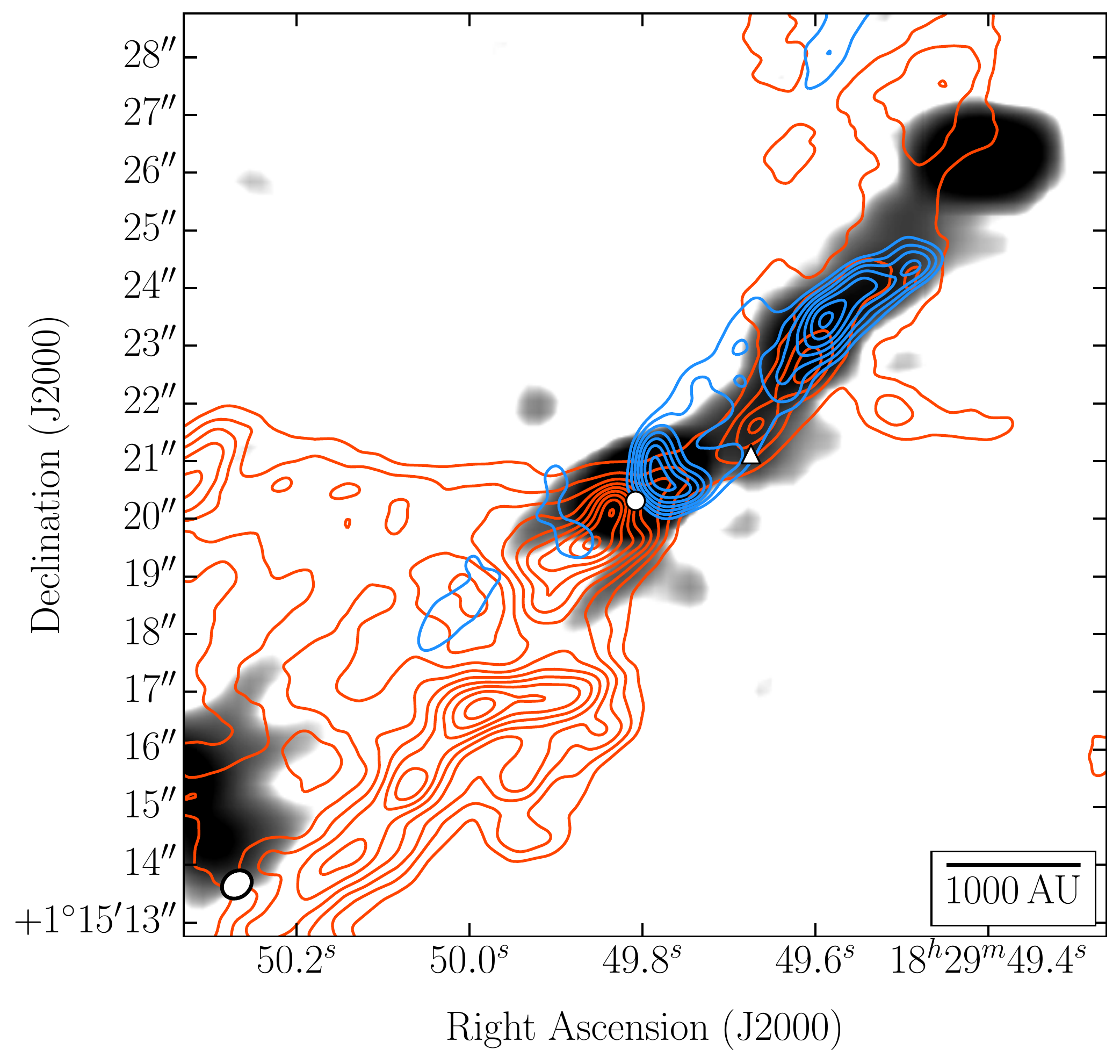}
\caption{\footnotesize   
Low-velocity red- and blueshifted \co{} from the ALMA data (red and blue contours, respectively), overlaid on the VLA 4\,cm continuum image (grayscale).  As in Figure \ref{fig:ALMA_VLA}, the image is centered on SMM1-a.  The CO velocity ranges are 2~to~15\,\kms{} (redshifted) and --20~to~--5\,\kms{} (blueshifted) relative to the \vlsr{} of SMM1 of approximately 9\,\kms{}.  
The contours are plotted at levels of 0.15, 0.25, 0.35, 0.45, 0.55, 0.65, 0.75, 0.85, 0.95 $\times$ the peaks of the redshifted (3.76\,\jybmkms{}) and blueshifted (4.16\,\jybmkms{}) moment 0 maps.
% ORIGINAL VLSR VALUES: 12 to 25 km/s (redshifted) and --10 to 5 km/s (blueshifted).
} 
\vspace{0.2in}
\label{fig:co_low} 
\end{figure}

The EHV CO corresponds well to what is seen toward other young protostellar sources with molecular jets, such as L1448C and HH 212 \citep{CFLee2015}.
The CO jet extends about 5$\arcsec$ ESE of SMM1-a (see Figure \ref{fig:ALMA_VLA}, right panel).  We also detect collimated, high-velocity emission toward the extreme SE of the image, just beyond the SE radio knot (about 12$\arcsec$ SE of SMM1-a; see Figure \ref{fig:ALMA_VLA}, left panel).  These two CO features are not perfectly collinear; furthermore, the four main radio knots are not all collinear.  \citet{RodriguezKamenetzky2016} suggest that the precession of the jet could cause this non-collinearity; they attribute the precession to an unresolved 3\,AU binary within SMM1-a and calculate a precession period of 20--30\,yr and a precession angle of 10$\degree$ (i.e., the angle between the central jet axis and the line of maximal deviation from that axis).  

In addition to the EHV CO jet, there is also wide-angle, \textit{low-velocity} redshifted CO emission to the SE of the source (see Figure \ref{fig:co_low}); this emission is consistent with the low-velocity CO reported in \citet[][Figure 27]{Hull2014}. 
The low-velocity CO along the walls of the cavity may have been accelerated outward by a low-velocity, wide-angle wind that is distinct from the EHV jet \citep[e.g.,][]{SantiagoGarcia2009, Arce2013}, or may be the result of shock entrainment by the central EHV jet \citep[e.g.,][]{Gueth1998, Gueth1999}.

Although no EHV CO emission is apparent from the blueshifted lobe of the outflow, Figure \ref{fig:co_low} shows low-velocity CO emission that is coincident with the ionized jet that extends in the NW direction.  The asymmetry of both the EHV CO and the ionized jets are puzzling, and will be investigated in more depth in a future paper.

\subsection{Characterizing the free-free emission}

\begin{figure} [hbt!]
\centering
\includegraphics[scale=0.485, clip, trim=0.6cm 0cm 0cm 0cm]{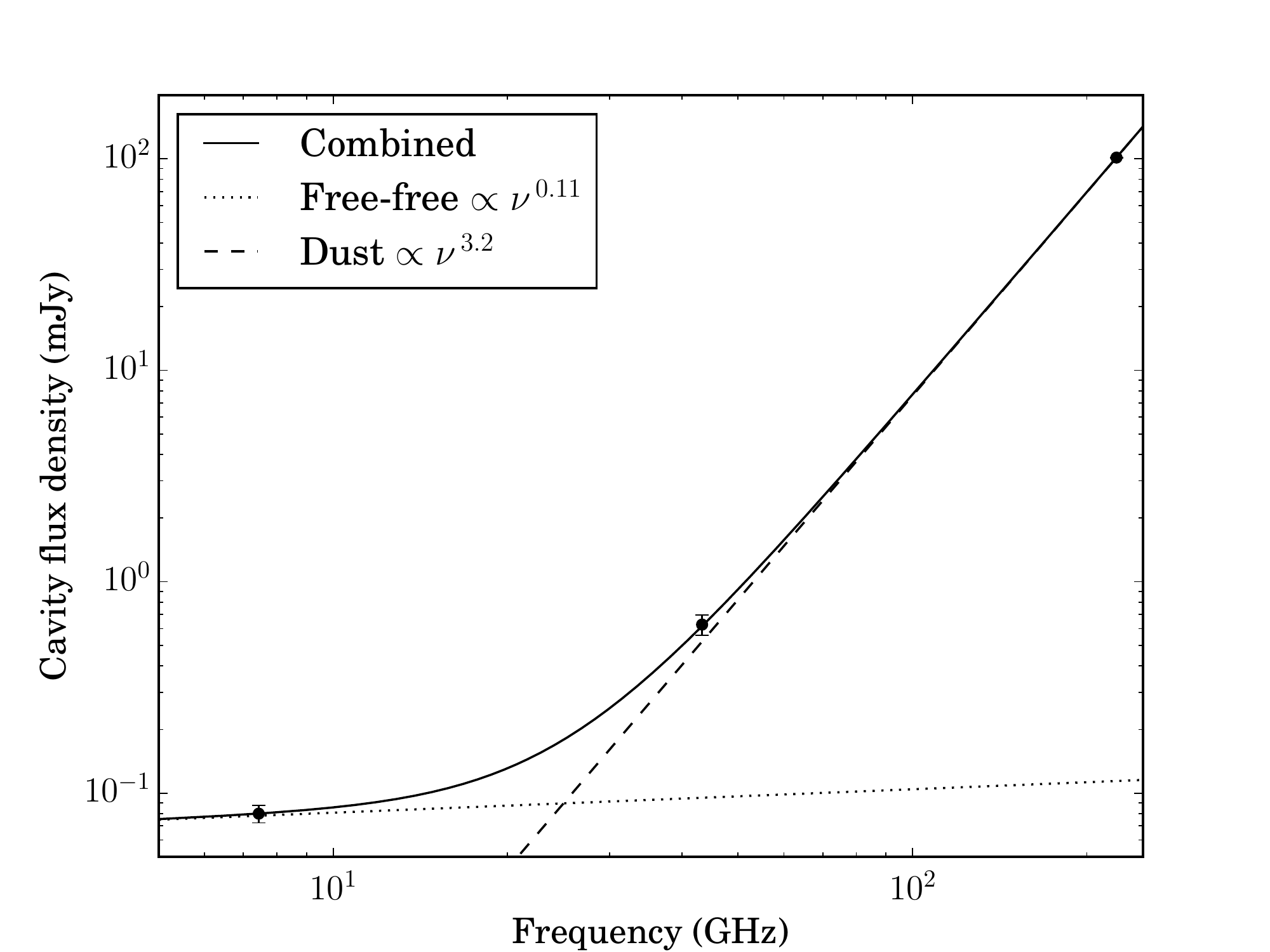}
\caption{\footnotesize   
Flux from the outflow cavity at 1.3\,mm (ALMA), 7\,mm (VLA, \citealt{Choi2009}), and 4\,cm (VLA).  The fitted curves include free-free (dotted) and a thermal dust (dashed) components.  The fluxes are the sum of the emission from the northern and southern sides of the cavity (see Footnote \ref{footnote:smoothing}).  The dust spectral index was fixed at a conservative value of 3.2; at 1.3\,mm we assume all emission is from dust.  The 4\,cm flux is dominated by free-free emission.
} 
\vspace{0.2in}
\label{fig:flux_fit} 
\end{figure}

The coincidence of the centimeter emission with the millimeter dust emission in the cavity is unexpected, as previous observations have shown free-free emission to be confined to protostellar jets. This begs the question: could the centimeter emission be from dust as well?  To address this question we calculated the spectral index of the emission from the cavity
%Using maps made with matching spatial frequencies (a.k.a. \textit{uv}-coverage; the C\,+\,X-band VLA data cover 1--309\,k$\lambda$; the combined ALMA data cover 9--980\,k$\lambda$), 
by fitting the fluxes from the 1.3\,mm data from ALMA, 7\,mm data from the VLA from \citet{Choi2009}, and the 4\,cm VLA data.\footnote{The three maps we used for the flux measurements sampled nearly identical spatial frequencies.  For consistency, before fitting the fluxes we smoothed all of the maps with a circular Gaussian with a FWHM of 1$\arcsec$.  The fluxes were calculated by summing the fluxes in two circular 1$\arcsec$ areas placed on the northern and southern cavity lobes.  The uncertainties in the fluxes plotted in Figure \ref{fig:flux_fit} are $ \sqrt{2} \,\, \times $ the rms noise levels in the maps, since two independent beams were used to measure the northern and southern fluxes. \label{footnote:smoothing}}  We assumed the 1.3\,mm emission was all from dust, and fixed the dust emission spectral index to a conservative value of 3.2---significantly lower than the typical spectral index ($\sim$\,3.5--4.0) of optically thin, micron-sized dust grains.  Even in this conservative case the 4\,cm flux is clearly dominated by free-free emission with a spectral index between the expected values of --0.1 and 0.6 (see Figure \ref{fig:flux_fit}).

\section{DISCUSSION}

Both shocks and UV ionization could cause the free-free emission seen in the cavity around the EHV jet.  UV ionization and shocks from winds of various speeds are expected to be common phenomena in YSOs \citep{Visser2012, Goicoechea2012}.  We explore three scenarios involving these two mechanisms:  
the ionization could be caused by (1) UV radiation produced by accretion of material onto the central protostar; (2) material from a precessing jet interacting with the cavity walls on a timescale shorter than the free-free cooling time; or (3) shocks from wide-angle winds.

\subsection{Cavity ionization by UV radiation}
\label{sec:UV}

The slow component of the outflow has clearly evacuated a cavity even wider than the one seen in free-free emission (see Figure \ref{fig:co_low}), which could allow UV radiation produced near the protostar to escape and ionize the cavity walls \citep{Visser2012}.  
We explore the possibility that UV radiation produced from the accretion of material onto the protostar could produce the free-free emission we see toward both the central source and the ionized cavity that lies $\sim$\,800\,AU from the central source. 

The rate of ionizing, Lyman-continuum photons required to sustain a given amount of free-free emission is as follows \citep{Rubin1968, MartinHernandez2003}:

\begin{equation}
N_{\textrm{Ly}}\,\,\textrm{(s$^{-1}$)} = 4.76 \times 10^{42} \,\, S_{\nu, \textrm{Jy}} \,\, \nu^{0.1}_{\textrm{GHz}} \,\, d_{\textrm{pc}}^{\,2} \,\, T_{e,\textrm{K}}^{\,-0.45} \,\,.
\label{eqn:UV}
\end{equation}

\noindent
The free-free emission from the central source and the cavity in SMM1 has an integrated flux density of $S_{\nu}$\,$\approx$\,2\,mJy at $\nu = 7.44$\,GHz. Assuming a distance to the source $d = 415$\,pc, an electron temperature $T_{e} = 10,000$\,K, and a flux of purely Ly-continuum photons with energies of 13.6\,eV, the Lyman continuum flux required to produce the free-free emission from SMM1 is $L_{\textrm{Ly-cont}} \approx 6.9 \times 10^{32}$\,\ergpersec{}. Based on results from \citet{France2014}, a classical T Tauri star's typical far-UV (FUV) luminosity, which is dominated by accretion, is between roughly $10^{30}$ and $10^{33}$\,\ergpersec{}. Furthermore, the results from \citeauthor{France2014} show that only $\sim$\,1\% of the FUV emission is in the Lyman continuum (13.6\,eV); the rest is in the Ly-alpha line (10.2\,eV).  Therefore, $L_{\textrm{FUV}}$ (dominated by Ly-alpha emission) $\sim$\,100 $\times$ $L_{\textrm{Ly-cont}}$.  Typical values of $L_{\textrm{FUV}}$ / $L_{\textrm{bol}}$ are in the range of $10^{-3} - 10^{-1}$. Scaling the 13.6\,eV luminosity of SMM1 to the full FUV luminosity, we find $L_{\textrm{FUV}}$ to be $\sim$\,10\,$L_\odot$.  $L_{\textrm{bol}}$ of SMM1 is 100\,$L_\odot$ \citep{Kristensen2012}, and thus $L_{\textrm{FUV}} / L_{\textrm{bol}} \sim 0.1$, consistent with what is seen toward classical T Tauri stars.\footnote{Given the FUV flux of SMM1 ($6.9 \times 10^{32}$\,\ergpersec{}), the radiation field at $\sim$\,800\,AU (the distance from the central source to the cavity) would be $38$\,\ergpersecpercmsq{}. This is approximately $2.4 \times 10^{4}$ times the interstellar FUV radiation field $G_0 = 1.6 \times 10^{-3}$\,\ergpersecpercmsq{} \citep{Habing1968}.}  

Note that Equation \ref{eqn:UV} does not take optical depth into account.  Assuming a constant volume density of $\sim$\,3000\,cm$^{-3}$ within the outflow cavity \citep{Visser2012}, the total column density out to 800\,AU is $\sim\,3.6 \times 10^{19}$\,cm$^{-2}$, or $\sim$\,0.02 $A_v$.  This corresponds to an optical depth in the UV of $\tau_{\textrm{UV}} \approx 0.06$, or an attenuation of only 6\% percent of the UV radiation \citep{Visser2012}.  For a cavity column density of $10^4$\,cm$^{-3}$, attenuation increases to $\sim$\,20\%.  Thus, if the cavity densities are $\lesssim$\,$10^4$\,cm$^{-3}$, we can safely ignore UV attenuation. 

However, even if the density within the cavity is higher than $10^4$\,cm$^{-3}$, SMM1 has a much larger mass, accretion rate, and FUV luminosity than even DF Tau, the brightest source in the sample from \citet{France2014}. It is therefore plausible that the free-free emission from both the central source and the cavity is a result of ionization by UV radiation created by the accretion shock.

\subsection{Cavity ionization by the precessing jet}

\begin{figure} [hbt!]
\centering
\includegraphics[scale=0.39, clip, trim=0cm 0cm 0cm 0cm]{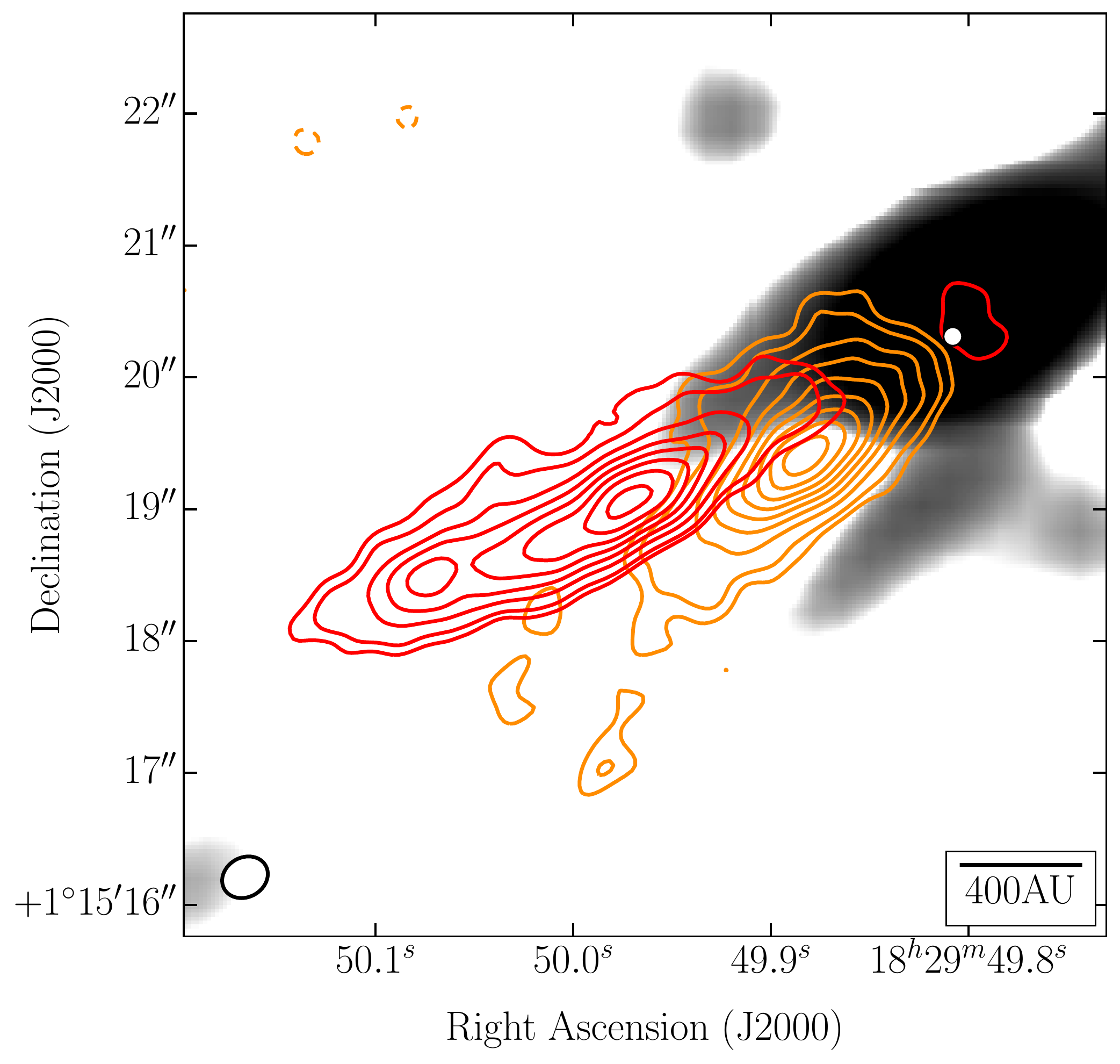}
\caption{\footnotesize   
The EHV \co{} jet imaged at two velocity ranges: 33--57\,\kms{} (orange) and 63--78\,\kms{} (red).  We attribute the differing orientations and physical extents of the emission to projection effects brought about by the jet precession.  
The CO contours are --3, 3, 4, 6, 8, 10, 12, 14, 16, 18 $\times$ the rms noise level in the moment 0 maps of 30\,\mjybmkms{}.
The grayscale is the 4\,cm VLA free-free emission, as plotted in Figure \ref{fig:ALMA_VLA} (right).
% ORIGINAL VLSR VALUES: (\vlsr{} = 43--67\,\kms{}), (\vlsr{} = 73--88\,\kms{})
} 
\vspace{0.2in}
\label{fig:co_multiple} 
\end{figure}

Figure \ref{fig:co_multiple} shows that the orientation and physical extent of the highest velocity CO emission (63--78\,\kms{}) are substantially different from those of the lower velocity emission (33--57\,\kms{}).  The differing orientations are evidence of the jet's $\sim$\,20\,yr precession through a $\sim$\,10$\degree$ angle \citep{RodriguezKamenetzky2016}; the slightly different emission extents and velocities 
are most likely the result of projection effects brought about by the precession.

While the morphology of the free-free cavity is roughly symmetric, there is significantly less dust emission from the northern flank of the cavity (i.e., the portion of the cavity $\sim$2$\arcsec$ directly to the east of the peak of SMM1-a).   %Considering that (sub)millimeter dust emission morphology can change on timescales of months \citep{Johnstone2013}, 
This suggests that the jet has interacted with the edge of the cavity. The energy injected into the cavity walls by these shocks is sufficient to produce the $\sim$\,0.1\,mJy of free-free emission that we see \citep[][Equation 25]{Curiel1987}.  

Assuming that the material from the precessing EHV outflow periodically impacts the cavity walls on a 20\,yr timescale, we calculate the free-free cooling time to test whether both sides of the cavity could remain ionized during the jet's precession period.  The recombination time $t_{\textrm{cool}} = \left(n_e \alpha\right)^{-1}$, where $n_e$ is the electron density and $\alpha$ is the recombination coefficient that depends on the electron temperature $T_e$ \citep[][Eqn. 5-14]{Spitzer1978}: 

\begin{equation}
\alpha \approx 4 \times 10^{-11} \, \left( \frac{T_e}{\textrm{K}} \right)^{-1/2}\,\textrm{cm}^3\,\textrm{s}^{-1}.
\end{equation}

\noindent
Therefore, from \citet[][Eqn. 6-11]{Spitzer1978}:

\begin{equation}
t_{\textrm{cool}}\,(\textrm{yr}) \approx  8 \, \left( \frac{T_e}{10^4\,\textrm{K}} \right)^{1/2} \left(\frac{n_e}{10^4\,\textrm{cm}^{-3}} \right)^{-1}\,.
\end{equation}

\noindent
For $T_e \sim 10^4$\,K (the typical temperature of an HII region) and $n_e \ga 10^4$\,cm$^{-3}$, the recombination timescale is within a factor of 2 of the precession period.  It is therefore plausible that shocks from the EHV jet interacting with the cavity wall could produce the free-free emission we observe.

\subsection{Cavity ionization by a wide-angle wind}

Finally, the ionization in the cavity could be caused by shocking of material from a (fast and undetected) wide-angle wind with the cavity walls.  
Fast ($\sim$\,100\,\kms{}), wide-angle winds such as X-winds \citep{Shu2000} or disk winds \citep{Konigl2000} are predicted to exist alongside collimated jets \citep{Shang2006, Shang2007}, and have been detected before in older, Class I protostars in a variety of atomic and ionized tracers \citep[e.g.,][]{Davis2002}. 
Evidence of shocks from winds in low-mass protostellar sources have been seen before \citep[e.g.,][]{Goicoechea2012, Kristensen2013a, Kristensen2013b}.

\citet{Kristensen2013b} argue that powerful shocks from fast, long-lived winds cannot be the sole cause of the ionization because those winds would disrupt the core in $\sim$\,1000 years, far less than the lifetime of a typical Class 0 protostar.  Rather, they argue that ionization could arise from one of two scenarios: (1) from winds that are episodic in time and direction, and thus fail to disrupt the core, or (2) a combination of a slow wind and UV radiation from the protostar itself.

\vspace{0.1in}
\section{CONCLUSION}

We have reported the highest resolution and highest sensitivity millimeter-wave observations to date of the intermediate-mass protostar Serpens SMM1.  The ALMA data show a one-sided, extremely high-velocity ($\sim$\,80\,\kms{}) \co{} jet powered flanked at the base by a wide-angle cavity seen in 4\,cm free-free emission detected by the VLA.  
This is the first time that ionization of an outflow cavity has been directly detected via free-free emission in a very young, embedded Class 0 protostellar source that is still powering a molecular jet.

The two most likely scenarios that could lead to the ionization of the outflow cavity are: (1) the cavity walls are ionized by UV photons escaping from the accreting protostellar source, or (2) material from the precessing molecular jet is impacting the walls and ionizing them via shocks.  We find both of these scenarios to be plausible---and indeed, the ionization may be caused by a combination of the two, as predicted in various studies.

If the main cause of the cavity's ionization is shocks from the precessing jet and the jet does indeed precess appreciably on a $\sim$\,20\,yr timescale, future ALMA and VLA observations could confirm whether the free-free emission, dust continuum, or the EHV CO emission change with time.  Furthermore, with the exquisite sensitivity of the VLA, it should be possible to detect ionized outflow cavities toward other bright, Class 0 protostellar sources, enhancing our knowledge of the radiation, outflow, and jet environments in the youngest, most embedded forming stars.

\acknowledgements

The authors acknowledge the anonymous referee, whose constructive comments improved this work substantially.
The authors also thank Ruud Visser for his careful reading of our paper and for his timely communication of a necessary correction.
C.L.H.H. acknowledges the invaluable data reduction assistance from Crystal Brogan and Bill Cotton.
The authors acknowledge Minho Choi for kindly supplying them with a FITS file of the 7\,mm VLA data for the purpose of calculating the spectral index of the emission from the outflow cavity.
J.M.G. acknowledges support from MICINN AYA2014-57369-C3-P and the MECD PRX15/00435 grants (Spain).
L.E.K. and M.M.D. acknowledge support from the Submillimeter Array (SMA) through the SMA postdoctoral fellowship.
M.M.D. acknowledges support from NASA ADAP grant NNX13AE54G. 
A.R.-K. and C.C.-G. acknowledge support by UNAM-DGAPA-PAPIIT grants numbers IA101214 and IA102816 and financial support from a MINCyT-CONACyT ME/13/47 grant, corresponding to the Bilateral Cooperation Program.
Z.-Y. L. is supported in part by NASA NNX14AB38G and NSF AST-1313083.
The National Radio Astronomy Observatory is a facility of the National Science Foundation operated under cooperative agreement by Associated Universities, Inc. 
This paper makes use of the following ALMA data: ADS/JAO.ALMA\#2013.1.00726.S. 
ALMA is a partnership of ESO (representing
   its member states), NSF (USA) and NINS (Japan), together with NRC
   (Canada), NSC and ASIAA (Taiwan), and KASI (Republic of Korea),
   in cooperation with the Republic of Chile. The Joint ALMA
   Observatory is operated by ESO, AUI/NRAO and NAOJ.
This research made use of APLpy, an open-source plotting package for Python hosted at \url{http://aplpy.github.com}.

%\bibliography{ms}
%\bibliographystyle{apj}

\end{document}